# The speed of cool soft pions

J. M. Martínez Resco [*],

M. A. Valle Basagoiti [†]

*Departamento de Física Teórica,*

*Universidad del País Vasco, Apartado 644, E-48080 Bilbao, Spain*

(September 30, 2018)

## Abstract

The speed of cool pions in the chiral limit is analytically computed at low temperature within the imaginary time formalism to two loop order. This evaluation shows a logarithmic dependence in the temperature where the scale within the logarithm is very large compared to the pion decay constant.
11.10.Wx, 12.39.Fe, 14.40.Aq

Typeset using REVTEX

---

[*]wtbmarej@lg.ehu.es

[†]wtpvabam@lg.ehu.es

The dynamics of pions in a thermal bath has been studied intensively in the last few years [1–6] because its absorptive and dispersive properties can determine the termalization processes ocurring in the aftermath of high-energy nuclear collisions.

In the limit of exact chiral symmetry, pions are true Goldstone bosons and as such, they travel at speed of light in the vacuum. At non zero temperature, however, relativistic invariance is lost, the thermal bath providing a priviliged rest frame and pions travel at less than the speed of light. Accordingly, the pion dispersion relation as a function of momentum must be modified. In chiral perturbation theory to leading order in low temperature, this modification first shows up at two loops, $\sim T^4/F_\pi^4$, but in a linear $\sigma$ model the effect already appears at one loop order, $\sim T^4/F_\pi^2 m_\sigma^2$ [5].

Only very recently [6], the speed of thermal pions has been explicitly computed. This has been obtained from the axial current two-point correlator in the real time formalism within chiral perturbation theory. There, some integrals had to be numerically computed due to its complexity. The final result shows a logarithmic dependence in the temperature,

$$v^2 = 1 - \frac{T^4}{27\, F_\pi^4} \log\left(\frac{\Lambda}{T}\right), \tag{1}$$

where the scale obtained is $\Lambda \approx 1.8$ GeV.

In this letter, we follow a rather different approach to the evaluation of the speed of the pion. We will work in the imaginary time formalism and use the background field method to compute the required piece of the effective action. As we shall see below, only diagrams having external zero momentum and energy are required. Thus, we take advantage of the methods developed by Arnold an Zhai [7] to analytically compute integrals associated with higher-loop vacuum bubbles of the partition function in gauge theories. We find a result which is in agreement with the one previously given by Toublan [6] except for the scale within the logarithm. Here we find $\Lambda \approx 4.3$ GeV. In spite of this difference, the behavior of the pion speed is not greatly modified in the range of temperature where this calculation can be trusted, $T < F_\pi$.

In the limit of exact chiral symmetry $SU(2)_L \times SU(2)_R$ which is spontaneously broken to $SU(2)_V$, the pion interaction at small energies is known from the effective chiral lagrangian with an increasing number of derivatives corresponding to an expansion in powers of momentum [8],

$$\mathcal{L}_{eff} = \frac{F_\pi^2}{4}\text{Tr}[\partial_\mu U \partial^\mu U^\dagger]$$
$$+ L_1(\text{Tr}[\partial_\mu U \partial^\mu U^\dagger])^2 + L_2 \text{Tr}[\partial_\mu U \partial_\nu U^\dagger]\text{Tr}[\partial^\mu U \partial^\nu U^\dagger] + \mathcal{L}_6 + \ldots \quad (2)$$

Fortunately, the explicit form of $\mathcal{L}_6$, giving the tree level counterterms to be used in a two loop computation, is not required because its contribution at tree level is independent of the temperature. A very efficient technique which organizes the graphs to be computed is provided by background field perturbation theory. Here, the $U$ field is parametrized by $U = \xi(x)h(x)\xi(x)$ where $\xi(x) = \exp(i\pi_{cl}/2F_\pi)$ is a classical backgroung field and $h(x) = \exp(i\tilde{\pi}/F_\pi)$ is a fluctuating quantum field. When the lagrangian is written in terms of $\xi$ and $h$ it becomes $\mathcal{L}(\tilde{\pi}, A, V)$. This lagrangian has an exact local chiral $SU(2)$ invariance implemented by the connection $V_\mu = (\xi^\dagger \partial_\mu \xi + \xi \partial_\mu \xi^\dagger)/2$. Such a field is not propagating and the remaining fields $\tilde{\pi}, A_\mu = (\xi^\dagger \partial_\mu \xi - \xi \partial_\mu \xi^\dagger)/2$, transform covariantly under $SU(2)$ transformations. Thus, the covariant derivative on quantum fluctuations is

$$D_\mu \tilde{\pi} = \partial_\mu \tilde{\pi} + [V_\mu, \tilde{\pi}]. \quad (3)$$

It is our aim here to compute the effective action $\Gamma[\pi_{cl}]$ in order to obtain the pion inverse propagator from two functional derivations. This can be computed from the following equality [9]

$$\Gamma[\pi_{cl}] = \tilde{\Gamma}[\tilde{\pi}_{cl}; A[\pi_{cl}], V[\pi_{cl}]]\Big|_{\tilde{\pi}_{cl}=0}, \quad (4)$$

where $\tilde{\Gamma}[\tilde{\pi}_{cl}; A[\pi_{cl}], V[\pi_{cl}]]$ is calculated restricting ourselves to diagrams with no external $\tilde{\pi}_{cl}$ lines. These diagrams are the one-particle-irreducible subset with only internal $\tilde{\pi}$ lines and external $V_\mu, A_\mu$ lines. Since the required piece of the effective action is quadratic in $\pi_{cl}$, and the expansion of background fieds is

$$A_\mu = \frac{i}{2F_\pi}\partial_\mu \pi_{cl} + \mathcal{O}(\pi_{cl}^3), \quad V_\mu = \frac{1}{8F_\pi^2}[\pi_{cl}, \partial_\mu \pi_{cl}] + \mathcal{O}(\pi_{cl}^4), \quad (5)$$

we only need to consider the first few terms of the expansion

$$\tilde{\Gamma}[0; A[\pi_{cl}], V[\pi_{cl}]] = \int d^4x V_\mu^a(x)\langle V_\mu^a(x)\rangle_{1PI}$$
$$+\frac{1}{2}\int\!\!\int d^4x\, d^4y\ A_\mu^a(x)\langle A_\mu^a(x)A_\nu^b(y)\rangle_{1PI}A_\nu^b(y) + \ldots \qquad (6)$$

Expanding $h$ up to order $1/F_\pi^4$ we get for $\mathcal{L}_2$

$$\mathcal{L}_2(\tilde{\pi}; A, V) = -F_\pi^2 \text{Tr}[A_\mu A^\mu] + \frac{1}{4}\text{Tr}[D_\mu\tilde{\pi}D^\mu\tilde{\pi}] - \frac{1}{4}\text{Tr}[[A_\mu, \tilde{\pi}]^2]$$
$$-iF_\pi\,\text{Tr}[\partial_\mu\tilde{\pi}A_\mu] + iF_\pi\text{Tr}[\tilde{\pi}[V, A]] + \frac{2i}{3F_\pi}\text{Tr}[(\vec{\tilde{\pi}}\vec{\tilde{\pi}}\partial_\mu\tilde{\pi} - \vec{\tilde{\pi}}\partial_\mu\vec{\tilde{\pi}}\tilde{\pi})A_\mu]$$
$$-\frac{1}{6F_\pi^2}\text{Tr}[(\vec{\tilde{\pi}}\vec{\tilde{\pi}})(V - A)\tilde{\pi}(V + A)\tilde{\pi}] + \frac{1}{6F_\pi^2}\text{Tr}[(\vec{\tilde{\pi}}\vec{\tilde{\pi}})^2(V - A)(V + A)]$$
$$-\frac{5i}{12F_\pi}\text{Tr}\left[(\vec{\tilde{\pi}}\vec{\tilde{\pi}})[V, A]\pi\right] + \frac{1}{6F_\pi^2}\text{Tr}\left[(\vec{\tilde{\pi}}\vec{\tilde{\pi}})[\pi, \partial_\mu\pi]V^\mu\right]. \qquad (7)$$

For $\mathcal{L}_4$ we write only the terms that will contribute at tree and one loop order,

$$\mathcal{L}_4 = L_1\left[-\frac{8}{F_\pi^2}\text{Tr}[A_\mu A^\mu]\text{Tr}[\partial_\mu\tilde{\pi}\partial^\mu\tilde{\pi}] - \frac{16}{F_\pi^2}\text{Tr}[A_\mu\partial^\mu\tilde{\pi}]\text{Tr}[A_\nu\partial^\nu\tilde{\pi}]\right]$$
$$+L_2\left[-\frac{8}{F_\pi^2}\text{Tr}[A_\mu A_\nu]\text{Tr}[\partial^\mu\tilde{\pi}\partial^\nu\tilde{\pi}]\right.$$
$$\left.-\frac{4}{F_\pi^2}\text{Tr}[A_\mu\partial_\nu\tilde{\pi} + A_\nu\partial_\mu\tilde{\pi}]\text{Tr}[A^\mu\partial^\nu\tilde{\pi} + A^\nu\partial^\mu\tilde{\pi}]\right] + \ldots \qquad (8)$$

The diagrams we have to compute from $\mathcal{L}_2$ are in Fig.1. The first gives a tree level contribution $\langle A_\mu^a A_\nu^b\rangle = -4F_\pi^2\delta^{ab}\delta_{\mu\nu}$. The fourth, fifth and the sixth diagrams vanish trivially due to the form of the vertices involved and we have $\langle V_\mu^a\rangle_{1PI} = 0$ up to two loops. The second, third and the seventh diagrams are necessarily proportional to $\delta^{\mu\nu}$. In consequence they give a contribution proportional to $P^\mu P^\nu \delta_{\mu\nu}$ which vanishes on shell. Then, only the last diagram at zero momentum remains to be computed. With the short-hand notation

$$\text{Tr}_K \longrightarrow \mu^{2\epsilon}\sum_{k_0=2\pi nT}\int\frac{d^{3-2\epsilon}k}{(2\pi)^{3-2\epsilon}}, \qquad (9)$$

it reads

$$P^\mu\langle A_\mu^a A_\nu^b\rangle P^\nu =$$
$$-\frac{1}{3!}\frac{64}{F_\pi^2}\delta^{ab}\,\text{Tr}_Q\text{Tr}_K\left[((PK)^2 + (PQ)^2 + PK\,PQ)\,\Delta(Q)\Delta(K)\Delta(K + Q)\right], \qquad (10)$$

where we have approximated the euclidean thermal propagator $\Delta(K+Q-P) \simeq \Delta(K+Q)$ because the required quadratic dependence in $P$ (see Eq. (15) below) is already present in the derivatives of $A[\pi_{cl}]$. Hence, it suffices to evaluate the sunset graph at zero external momentum. Some details for this calculation are given in the Appendix. The end result for the sunset contribution takes the form at $P^2 = 0$,

$$\frac{1}{4F_\pi^2} P^\mu \langle A_\mu^a A_\nu^b \rangle P^\mu =$$
$$\frac{1}{F_\pi^4} \delta^{ab} p^2 \left( -\frac{T^4}{54} \frac{1}{\epsilon} + \frac{\gamma_E T^4}{81} - \frac{2T^4}{27} \log(\frac{\mu}{T}) - \frac{46 T^4}{1215} + \frac{T^4}{81} \log(16\pi) \right.$$
$$\left. + \frac{80 T^4}{27} \zeta'(-3) + \frac{8 T^4}{9} \zeta'(-1) + \frac{20 T^4}{9 \pi^4} \zeta'(4) + \mathcal{O}(\epsilon) \right), \quad (11)$$

where $p$ denotes the three momentum. The only contribution from $\mathcal{L}_4$ is (see Fig.2),

$$\frac{1}{4F_\pi^2} P^\mu \langle A_\mu^a A_\nu^b \rangle P^\nu = -\frac{1}{4F_\pi^2} \frac{1}{2!} \frac{256 \, \delta^{ab}}{F_\pi^2} (L_1 + 2L_2) \, \mathrm{Tr}_Q (PQ)^2 \Delta(Q)$$
$$= -\frac{64\pi^2}{45} \frac{p^2 T^4}{F_\pi^4} (L_1 + 2L_2) \left[ 1 + \epsilon \left( 240 \, \zeta'(-3) - \gamma_E + \log(\frac{\mu^2}{4\pi T^2}) + \frac{13}{6} \right) + \mathcal{O}(\epsilon^2) \right]. \quad (12)$$

The renormalization at finite temperature is the same as for $T = 0$. This means that the use of the renormalization prescription [8]

$$L_i = L_i^r(\mu) - \frac{\gamma_i}{32\pi^2} \left( \frac{1}{\epsilon} + \log(4\pi) + 1 - \gamma_E \right), \quad (13)$$

with $\gamma_1 = 1/12$ and $\gamma_2 = 1/6$ for $SU(2)$ must cancel the pole terms. Indeed, putting together the contributions coming from $\mathcal{L}_2$ and $\mathcal{L}_4$ the poles cancel as they should.

Now we can easily obtain the inverse propagator

$$\Delta_{ab}^{-1}(x,y) = \left. \frac{\delta^2 \Gamma[\pi_{cl}]}{\delta \pi_{cl}^a(x) \delta \pi_{cl}^b(y)} \right|_{\pi_{cl}=0} = -\frac{1}{4F_\pi^2} \partial_x^\mu \partial_y^\nu \langle A_\mu^a(x) A_\nu^b(y) \rangle_{1PI}. \quad (14)$$

In momentum space $\Delta_{ab}^{-1}(P) = \delta^{ab} \left( \Delta_0^{-1}(P) + \Sigma(P) \right)$. Thus, we can read the dispersion relation from the retarded proper self-energy $\Sigma^{\mathrm{ret}}(\omega, p)$ which can be obtained from the euclidean counterpart by analytic continuation as $\Sigma^{\mathrm{ret}}(\omega, p) = \Sigma(p_0 = -i\omega + 0^+, p)$. If the retarded inverse propagator has a zero located at $\omega = \omega(p) - i\gamma(p)$ with $\gamma(p) > 0$ we get

$$v^2 = 1 + \frac{\text{Re}\Sigma^{\text{ret}}(\omega = p)}{p^2}, \tag{15}$$

$$\gamma(p) = -\frac{\text{Im}\Sigma^{\text{ret}}(\omega = p)}{2p}. \tag{16}$$

Then the speed of the pions in the chiral limit is

$$v^2 = 1 - \frac{T^4}{F_\pi^4}\left(\frac{1}{27}\log\left(\frac{\mu}{T}\right) + \frac{64\pi^2}{45}(L_1^r(\mu) + 2L_2^r(\mu))\right.$$
$$\left. - \frac{101}{4860} + \frac{2\gamma_E}{81} - \frac{1}{81}\log(16\pi) - \frac{8}{9}\zeta'(-1) - \frac{200}{27}\zeta'(-3) - \frac{20}{9\pi^4}\zeta'(4)\right) + \mathcal{O}(p). \tag{17}$$

This is the main result of this letter. The scale $\mu$ enters the calculation in such a way that, had we used a different scale $\mu'$ but kept the speed invariant, we would have had

$$L_i^r(\mu') = L_i^r(\mu) - \frac{\gamma_i}{16\pi^2}\log\left(\frac{\mu'}{\mu}\right), \tag{18}$$

which is the correct result for the running chiral coefficients. At the scale of the mass of the pion, $\mu = m_\pi$, we take [11,6] $32\pi^2(L_1^r + 2L_2^r) \simeq 1.66$ to yield

$$v^2 = 1 - \frac{T^4}{27F_\pi^4}\log\left(\frac{\Lambda}{T}\right), \tag{19}$$

with $\Lambda \simeq 4300$ MeV. Notice that we have obtained an analytic expression

$$-\frac{101}{4860} + \frac{2\gamma_E}{81} - \frac{1}{81}\log(16\pi) - \frac{8}{9}\zeta'(-1) - \frac{200}{27}\zeta'(-3) - \frac{20}{9\pi^4}\zeta'(4) \simeq -0.0538, \tag{20}$$

for the constant in Eq. (17). This constant was computed by numerical integration in [6], where the value $-0.0202$ is given. This discrepancy explains the difference in the scale $\Lambda$ from 1.8 GeV to 4.3 GeV. However, this difference between logarithmic scales has a moderate impact on the quantitative behavior of the pion speed in the range of temperature where this calculation can be trusted. Thus, the change produced in $v^2$ is less than 6% up to $T \sim 100$ MeV.

The pion damping rate can also be computed easily within the imaginary time formalism at leading order. The result is $\gamma(p) = p^2 T^3 \log(T/p)/(18\pi F_\pi^4) + \mathcal{O}(p^3)$ in agreement with the one previously reported in [2,10].

How do extend these results to the case of non-zero quark masses? In principle, this question could be answered by considering the new graphs coming from the breaking part of the effective Lagrangian $\mathcal{L}_{breaking} = F_\pi^2/4 \operatorname{Tr}[\chi U^\dagger + \chi^\dagger U] + \ldots$, where $\xi = 2Bm$, with $m$ the quark mass matrix. However, the use of background field perturbation theory which in the chiral limit greatly reduces the number of graphs to be computed, does not seem to simplify the computation now, since the Lagrangian $\mathcal{L}(\tilde{\pi}, \xi, \chi)$ obtained after the splitting of $U$ has lost its (local) invariance under chiral transformations. Moreover, the integrals involved are complicated by the pion mass, making their analytical computation more difficult. We hope to address this question in the future.

## Acknowledgements

Discussions with M. A. Goñi and J. L. Mañes are gratefully acknowledged. J.M.M. is supported by a UPV/EHU grant. This work has been also partially supported by CICYT under project No. AEN96-1668 and the University of the Basque Country grant UPV-EHU 063.310-EB225/95.

**APPENDIX A:**

To do our calculations we follow Arnold and Zhai [7] which allows us to evaluate the sum-integrals analytically. We can rewritte Eq. (10) (without the factors in front of it) as :

$$\frac{1}{4} \operatorname{Tr}_Q \Delta(Q)[P^\mu P^\nu \Pi^{\mu\nu}(Q) + 3(PQ)^2 \Pi(Q) + 2P^2 \operatorname{Tr}_K \Delta(K)], \tag{A1}$$

where

$$\Pi^{\mu\nu}(Q) = \operatorname{Tr}_Q[(2K+Q)^\mu (2K+Q)^\nu \Delta(K) \Delta(K+Q)] - 2\,\delta^{\mu\nu} \operatorname{Tr}_K \Delta(K), \tag{A2}$$

and

$$\Pi(Q) = \operatorname{Tr}_K \Delta(K) \Delta(K+Q). \tag{A3}$$

We start with the term proportional to $\Pi(Q)$. As the result is independent of the direction of $\vec{p}$ we average over its angles in $d-1$ dimensions,

$$\operatorname{Tr}_Q (p_0 q_0 + \vec{p}\vec{q})^2 \Delta(Q) \Pi(Q) = \left(p_0^2 - \frac{p^2}{d-1}\right) \operatorname{Tr}_Q q_0^2 \,\Pi(Q)\,\Delta(Q) + \frac{p^2}{(d-1)} \operatorname{Tr}_Q \Pi(Q). \tag{A4}$$

Following [7] we split $\Pi(Q)$ into two terms

$$\Pi(Q) = \Pi^{(0)}(Q) + \Pi^{(T)}(Q), \tag{A5}$$

where $\Pi^{(0)}(Q)$ is the zero-temperature contribution and $\Pi^{(T)}(Q) = \tilde{\Pi}^{(T)}(Q) + \Pi^{UV}(Q)$ is the temperature dependent piece. Although $\Pi^{(T)}(Q)$ is UV finite, its large-$Q$ asymptotic behaviour $\Pi^{UV}(Q) \sim 1/Q^2$, gives rise to a UV divergence in $\operatorname{Tr}_Q q_0^2 \,\Pi^{(T)}(Q)\,\Delta(Q)$. Thus, it is useful to evaluate separately the three contributions,

$$\operatorname{Tr}_Q q_0^2 \Delta(Q) \tilde{\Pi}^{(T)}(Q) = \frac{T^4}{1080}(7 + 3\gamma_E + 1440\,\zeta'(-3) + 180\,\zeta'(-1)) + \mathcal{O}(\epsilon), \tag{A6}$$

$$\operatorname{Tr}_Q q_0^2 \Delta(Q) \tilde{\Pi}^{UV}(Q) = -\frac{19 T^4}{2160} - \frac{T^4}{720}\frac{1}{\epsilon} - \frac{T^4 \gamma_E}{360} - \frac{T^4}{360}\log(\frac{\mu^2}{2\,T^2}) + \frac{T^4}{4\,\pi^4}\zeta'(4) + \mathcal{O}(\epsilon), \tag{A7}$$

$$\operatorname{Tr}_Q q_0^2 \Delta(Q) \Pi^{(0)}(Q) = -\frac{T^4}{480}\frac{1}{\epsilon} - \frac{T^4}{80} + \frac{T^4 \gamma_E}{240} + \frac{T^4}{240}\log(4\,\pi)$$
$$- T^4\,\zeta'(-3) - \frac{T^4}{120}\log(\frac{\mu}{T}) + \mathcal{O}(\epsilon). \tag{A8}$$

The remaining integral in Eq. (A4) gives

$$\text{Tr}_Q \Pi(Q) = \left(\text{Tr}_K \frac{1}{K^2}\right)^2 = \frac{T^4}{144} + \mathcal{O}(\epsilon). \tag{A9}$$

Now we have to evaluate $\text{Tr}_Q \Delta(Q) P^\mu P^\nu \Pi^{\mu\nu}(Q)$. We can use the orthogonality property $Q^\mu \Pi^{\mu\nu}(Q) = 0$ to transform it into (after the average over the angles of $\vec{p}$)

$$\text{Tr}_Q P^\mu P^\nu \Pi^{\mu\nu}(Q)\Delta(Q) = -\frac{5}{432}p^2 T^4 + \left(p_0^2 - \frac{p^2}{(d-1)}\right) \text{Tr}_Q \Pi^{00} \Delta(Q), \tag{A10}$$

where has been used $\text{Tr}_Q \Delta(Q) = T^2/12$. Following the same steps as for the case of $\Pi(Q)$ we get:

$$\text{Tr}_Q \tilde{\Pi}_T^{00} \Delta(Q) = -\frac{T^4}{1080}(1 - 21\gamma_E - 2880\zeta'(-3) - 540\zeta'(-1)) + \mathcal{O}(\epsilon), \tag{A11}$$

$$\text{Tr}_Q \Pi_{UV}^{00} \Delta(Q) = -\frac{7T^4}{720}\frac{1}{\epsilon} - \frac{7\gamma_E T^4}{360} + \frac{17T^4}{2160} + \frac{7T^4}{4\pi^4}\zeta'(4) - \frac{7T^4}{360}\log(\frac{\mu^2}{2T^2}) + \mathcal{O}(\epsilon), \tag{A12}$$

$$\text{Tr}_Q \Pi_0^{00} \Delta(Q) = -\frac{T^4}{1440}\frac{1}{\epsilon} - \frac{T^4}{3}\zeta'(-3) + \frac{\gamma_E T^4}{720} - \frac{11T^4}{2160} - \frac{T^4}{720}\log(\frac{\mu^2}{4\pi T^2}) + \mathcal{O}(\epsilon). \tag{A13}$$

Adding up all contributions we finally obtain the result given in Eq. (11).

The contribution from the counterterms is proportional to $\text{Tr}_Q (PQ)^2 \Delta(Q)$. This is required up to order $\epsilon$ because of the pole in $(L_1 + 2L_2)$,

$$\text{Tr}_Q (PQ)^2 \Delta(Q) = \left(p_0^2 - \frac{p^2}{(d-1)}\right) \text{Tr}_Q \left(\frac{q_0^2}{Q^2}\right) + \frac{p^2}{(d-1)} \text{Tr}_Q 1, \tag{A14}$$

where the last term is zero in dimensional regularization. The other term gives

$$\text{Tr}_Q (PQ)^2 \Delta(Q) = \frac{4\pi^2 p^2}{90}\left(1 + \epsilon\left(240\zeta'(-3) - \gamma_E + \log(\frac{\mu^2}{4\pi T^2}) + \frac{13}{6}\right)\right). \tag{A15}$$

FIGURES

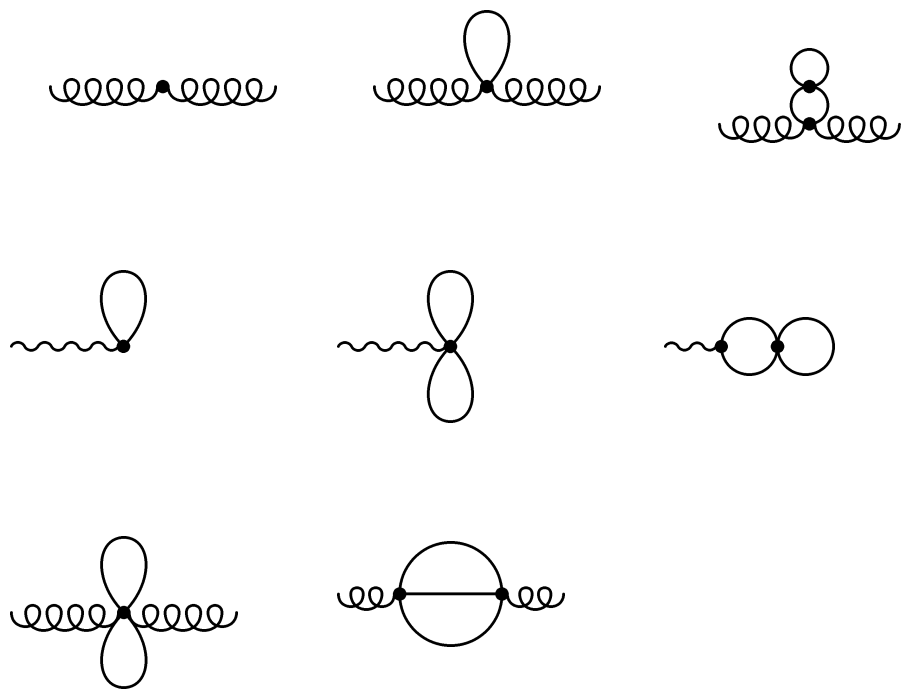

FIG. 1. Diagrams contributing from $\mathcal{L}_2$. The curly lines are associatted with the background field A, the wiggly lines with V, and the plain ones with the quantum fluctuations.

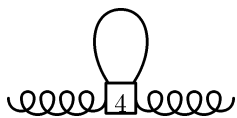

FIG. 2. Diagram contributing from $\mathcal{L}_4$.